\documentstyle[preprint,aps]{revtex}
\newcommand{\gsimeq}{\displaystyle{\mathop{>}_{\sim}}}
\newcommand{\lsimeq}{\displaystyle{\mathop{<}_{\sim}}}
\begin{document}
\draft
\preprint{\vbox{\hfill TSTC9601 \today}}
\title{
Wave Chaos in Quantum Billiards 
with Small but Finite-Size Scatterer
}
\author{
T. Shigehara
}
\address{
Computer Centre, University of Tokyo, Yayoi, 
Bunkyo-ku, Tokyo 113, Japan 
}
\author{
Taksu Cheon
}
\address{
Department of Physics, Hosei University, Fujimi, 
Chiyoda-ku, Tokyo 102, Japan
}
\date{January 17, 1996}
\maketitle
\begin{abstract}
We study the low energy quantum spectra 
of two-dimensional rectangular billiards 
with a small but finite-size scatterer inside. 
We start by examining the
spectral properties of billiards 
with a single pointlike scatterer. 
The problem is formulated in terms of self-adjoint extension
theory of functional analysis. 
The condition for the appearance of so-called wave chaos 
is clarified. 
We then relate the pointlike scatterer to 
a finite-size scatterer through the appropriate
truncation of basis. 
We show that the signature of wave chaos 
in low energy states
is most prominent 
when the scatterer is weakly attractive. 
As an illustration, numerical results of a rectangular 
billiard with a small rectangular scatterer inside are 
exhibited. 
\end{abstract}
\pacs{5.45.+b, 3.65.-w}
%
%\vfill{\ \\
%PACS Nos: 3.65.-w, 5.45.+b, 11.10.Gh\\
%-----------------------------------------------------\\
%email contact: takaomi@cc.u-tokyo.ac.jp, cheon@fujimi.hosei.ac.jp}
%
%\narrowtext
%\twocolumn
%\newpage
%
\section{Introduction}
%
%\paragraph*{}
The two-dimensional billiard is an appropriate tool 
for examining the generic features 
of dynamical systems because of 
the wide range of dynamical 
behaviors going from the most regular (integrable) to the 
most irregular (chaotic) depending on the geometry 
of its boundary. 
Although no mathematical proof exists, 
it is widely believed that fingerprints of the 
regular or irregular nature of the classical motion 
can be found in statistical properties of the quantum spectrum
both in 
energy levels and wavefunctions. 
The integrable systems such as the circular, elliptic and 
rectangular billiards obey the Poisson statistics \cite{BT77,CCG85}, 
while the predictions of the Gaussian orthogonal ensembles 
describe the chaotic systems such as Sinai's billiard and 
Bunimovich's stadium \cite{M67,B89}. 

Besides these extreme classes, there is an intermediate category  
called {\em pseudointegrable} \cite{RB81,Z92} 
or {\em almost integrable} \cite{ZK75} system. 
The typical examples are staircase billiards \cite{CC89}, 
rational polygons \cite{RB81,BJ90,SS93,SSSSSZ94}
and the billiards with pointlike scatterers 
inside an integrable one \cite{S90,CMSY91,SZ91,SYCM93,S94,CS95}. 
The nature of classical motion in pseudointegrable systems 
can be considered as integrable in the sense that 
unstable trajectories are of measure zero. 
However, several numerical studies have revealed that
under a certain condition,
the quantization induces the chaotic energy spectra,
which can be regarded as
a counterexample for the correspondence 
between the energy spectra and the underlying classical motion. 
This phenomenon is named {\em wave chaos} 
because its origin is the wave-like nature of quantum motion.
For the billiards with pointlike scatterers, 
this phenomenon has been understood in terms
of the quantum breaking of the classical 
scale invariance \cite{CS95}.
However, the notion of the pointlike scatterer is 
a mathematical abstraction whose relevance to the 
physical system is far from straightforward.
It is therefore highly desirable to show that 
the nature of wave chaos 
remains intact for the case of finite-size scatterer.
This is the prime motivation of the work we report in this paper.
We restrict ourselves to considering the billiards 
with a small but finite-size scatterer inside an integrable one. 
It is shown that, even if the scatterer has a substantial size,  
the signature of wave chaos are observable under a certain condition. 
The condition is consistent with a previous numerical observation 
in Sinai's billiard, which indicates that there exists no signature 
of wave chaos when the scatterer is repulsive.  

The paper is organized as follows. 
We start by examining the spectral properties of 
pseudointegrable billiards with a pointlike scatterer 
in Secs.II and III. In the former section, 
we give an accurate description 
of a pointlike scatterer, based on the self-adjoint extension theory  
in functional analysis. A special emphasis is laid on  
the distinction between symmetry and self-adjointness. 
We give an intuitive explanation of
the necessity of the self-adjointness for quantum mechanical Hamiltonians. 
Applying a general prescription for extending a symmetric operator 
to a self-adjoint one, 
we derive the Green's function for the billiards 
with a single pointlike scatterer. 
A renormalized coupling constant of the scatterer 
is defined in a natural manner, 
although its physical meaning is somewhat unclear at this stage. 
Based on the formulation in Sec.II, we discuss the quantum spectra 
of the billiards with a pointlike scatterer in Sec.III. 
The general 
condition for the appearance of wave chaos is clarified. 
In Sec.IV, we consider a Dirac's delta function potential 
with a truncated basis. 
Although it is not well-defined with a full basis 
in two-dimensional billiards, 
we can deduce certain physical contents from the truncated
system,
which serve as the basis for the ensuing discussions. 
The findings in Secs.III and IV are applied to the cases 
of a small but finite-size scatterer in Sec.V. 
It is shown that the wave chaos is manifest 
at low energy with attractive potentials. 
As an illustration, we give a numerical example for a rectangular 
billiard with a small rectangular scatterer inside. 
We also show a proper procedure for the zero-size limit
of the finite-size scatterer, which is consistent 
with the self-adjointness of the Hamiltonian. 
This clarifies the physical meaning of a renormalized coupling 
constant defined in Sec.II. 
We give the conclusions in Sec.VI. 
\section{Quantum Mechanical Formulation of Pseudointegrable Billiards 
         with a Pointlike Scatterer}
In spite of the apparent simplicity, 
careful treatments are required  for the definition of 
quantum billiards with a pointlike scatterer inside. 
Several methods are known for this purpose \cite{AGHH88}.  
Here, we adopt the one based on the
self-adjoint extension theory in functional analysis \cite{RS75}. 
We briefly summarize the formalism
stressing the necessary points for the 
discussions in following sections. 
We also mention the physical reasons 
why quantum mechanics requires the self-adjointness 
for Hamiltonians. 

Let us consider an integrable billiard of area $S$ 
with the Dirichlet boundary condition such that 
wavefunctions vanish on the boundary.  
The eigenvalues and corresponding eigenfunctions are determined by 
the stationary Schr\"{o}dinger equation; 
\begin{equation}
\label{1}
H_0 \varphi_n (\vec{x}) \equiv -\frac{1}{2M}\Delta \varphi_n (\vec{x}) 
= E_{n} \varphi_n (\vec{x})   (n = 1, 2, 3, \cdots), 
\end{equation}
where $\vec x$ is the coordinate vector in two-dimensional space, 
and $M$ is the mass of a particle moving in the billiard. 
We adopt natural units throughout this paper; $\hbar = 1, c=1$.  
This leaves a single independent unit among 
mass, energy and length. 
The domain of $H_0$ is $D(H_0)=H^2 (S) \cap H^1_0 (S)$ in terms of 
the Sobolev spaces. 
(We denote the domain of the billiard by the same symbol as the 
area unless there is danger of confusion.) 
In the following, we assume that the unperturbed billiard 
has no degeneracy. 
The Green's function of this system is given by 
\begin{equation}
\label{2}
G^{(0)}(\vec{x},\vec{y};z)=\sum_{n=1}^{\infty} 
\frac{\varphi_{n}(\vec{x}) \varphi_{n}(\vec{y})}
     {z-E_n}
\end{equation}
where $z$ is an energy variable. 
Suppose that a pointlike scatterer is placed at $\vec{x}=\vec{x}_0$ 
inside the billiard. 
Following the self-adjoint extension theory in functional analysis, 
the eigenvalues and the corresponding eigenfunctions 
of the perturbed system are calculated as follows.  

The first step for this purpose is  
to remove the relevant scattering point $\vec{x}_{0}$ 
by restricting $H_0$ to $T=H_0 \lceil D(T)$ with the domain 
\begin{equation}
\label{3}
D(T)= \{ \varphi(\vec{x}) \in D(H_0) \mid \varphi(\vec{x}_0)=0 \}.
\end{equation}
By using integration by parts, it is easy to see 
that $T$ is {\em symmetric (Hermitian)}. 
However, the domain of $T^*$, the adjoint operator of $T$, 
is not identical to $D(T)$ and indeed larger than $D(T)$; 
\begin{equation}
\label{4}
D(T^*)=D(T) \oplus Ran(T-\bar{\lambda})^{\perp} \oplus 
Ran(T-\lambda)^{\perp},  
\end{equation}
where the orthogonal complement to the range of $T-\lambda$ 
corresponds to the deficiency subspace of $T$ given by \cite{Z80}
\begin{equation}
\label{5}
Ran(T-\lambda)^{\perp}=\{ \varphi (\vec{x}) \mid 
\varphi (\vec{x})=
cG^{(0)}(\vec{x},\vec{x}_0,\bar{\lambda}),c \in {\bf C}  \} 
\end{equation}
with an arbitrary complex number $\lambda$ ($Im \lambda \neq 0)$. 
It follows from $D(T^*) \neq D(T)$ that 
$T$ is not {\em self-adjoint}. 
A symmetric operator is self-adjoint if and only if 
$D(T^*) = D(T)$. 

The two following facts serve to 
understand the reason for imposing self-adjointness on 
quantum mechanical Hamiltonians \cite{F67,S77}. 
One of them is based on the assertion
\begin{equation}
\label{6}
Ker(T^*-\bar{\lambda})=\overline{Ran(T-\lambda)}^{\perp}. 
\end{equation}
This means that $T^*$ has a complex eigenvalue 
$\bar{\lambda}$ with eigenspace 
$\overline{Ran(T-\lambda)}^{\perp}$, 
namely, the closure of $Ran(T-\lambda)^{\perp}$. 
If and only if $T$ is self-adjoint, 
$Ran(T-\lambda)^{\perp}= \emptyset$, {\em  i.e.},  
$Ran(T-\lambda)$ is the entire Hilbert space $X=L^2 (S)$, 
the set of all square-integrable functions over $S$.  
This is one of the reasons why quantum mechanics, 
which does not allow complex eigenvalues for observables, 
requires self-adjoint Hamiltonians. 
The second fact is closely related to the unitarity of time-evolution 
operators. 
The time-evolution operator $U(t)$ is constructed from $T$ as follows; 
\begin{equation}
\label{7}
U(t) \equiv e^{-iTt} 
= \lim_{n \rightarrow \infty}(1+\frac{iTt}{n})^{-n}
= \lim_{n \rightarrow \infty} \{ (\frac{it}{n})^{-n}
 (T-\frac{in}{t})^{-n}\}. 
\end{equation}
It is easy to see that, for any symmetric $T$, 
an operator $(T-\lambda)^{-1}$ 
with pure imaginary $\lambda \neq 0$ is defined 
from $Ran(T-\lambda)$ to $D(T)$ 
with $\|(T-\lambda)^{-1} \varphi\| \leq \| \varphi\| / |Im \lambda | $ 
for any $\varphi \in Ran(T-\lambda)$. 
If $Ran(T-\lambda)$ is not all of $X$, 
then in general,  $(T-\lambda)^{-n}$ can be defined 
on smaller and smaller spaces as $n$ increases. 
This leads us to a natural condition 
$Ran(T-\lambda)=X$, implying that time-evolution operators 
which preserve probability can be constructed only from   
self-adjoint operators.  

The above considerations indicate that 
we have to extend $D(T)$, or equivalently, restrict $D(T^*)$, 
in order to construct a self-adjoint operator from $T=H_0 \lceil D(T)$. 
This is indeed possible since 
$dim Ran(T-\bar{\lambda})^{\perp} = dim Ran(T-\lambda)^{\perp}$. 
Because both the deficiency subspaces are one-dimensional,  
a one-parameter family of self-adjoint extensions of $T$ exist. 
According to a general prescription for extending a symmetric operator 
to self-adjoint ones \cite{N32}, 
All the self-adjoint extensions of $T$ 
are given by $H_{\theta}=H_{0}$ $(0 \leq \theta < 2\pi)$ with the domain 
\begin{eqnarray}
\label{8}
D(H_{\theta})& = & \{ \psi(\vec{x}) \mid \psi(\vec{x})= \varphi(\vec{x}) 
+cG^{(0)}(\vec{x},\vec{x}_{0};\lambda) 
-ce^{i\theta}G^{(0)}(\vec{x},\vec{x}_{0};\bar{\lambda}); \nonumber \\
& & \varphi(\vec{x}) \in D(T), 
G^{(0)}(\vec{x},\vec{x}_{0};\lambda) \in Ran(T-\bar{\lambda})^{\perp}, 
G^{(0)}(\vec{x},\vec{x}_{0};\bar{\lambda}) \in Ran(T-\lambda)^{\perp}, 
\nonumber \\
& & c \in {\bf C} \}.
\end{eqnarray}
Eq.(\ref{6}) with Eq.(\ref{5}) indicates that 
the operation of $H_{\theta}$ on $\psi(\vec{x})$ is given by 
\begin{equation}
\label{9}
(H_{\theta}\psi)(\vec{x}) = (H_{0}\varphi)(\vec{x}) 
+ \lambda cG^{(0)}(\vec{x},\vec{x}_{0};\lambda)
- \bar{\lambda} ce^{i\theta}G^{(0)}(\vec{x},\vec{x}_{0};\bar{\lambda}).
\end{equation}
The operator $H_{\theta}$ is regarded as the Hamiltonian 
of the perturbed system with a pointlike scatterer. 
Notice that, although we have $H_{\theta}=H_{0}$ on $D(T)$, 
$D(H_{\theta})$ is substantially larger than $D(T)$; 
$D(T) \subset D(H_{\theta}) \subset D(T^*)$. 
We recognize from the second and third terms 
of $\psi (\vec{x})$ in Eq.(\ref{8}) that 
the appropriate boundary condition around the scatterer is 
specified after the extension. 
It will be shown later that the value of 
$\theta$ is related to the strength of the pointlike scatterer. 

With the aid of the resolvent equation, we can write down
an equation for 
the Green's function of the system 
with the Hamiltonian $H_{\theta}$ as 
\begin{equation}
\label{10}
G_{\theta}(\vec{x},\vec{y};z)=G^{(0)}(\vec{x},\vec{y};z)+
G^{(0)}(\vec{x},\vec{x}_{0};z)
T_{\theta}(z)
G^{(0)}(\vec{x}_{0},\vec{y};z),
\end{equation}
where the transition matrix (T-matrix) $T_{\theta}(z)$ is calculated through 
\begin{equation}
\label{11}
T_{\theta}(z) =\frac{1-e^{i\theta}}
{(z-\lambda)\int
G^{(0)}(\vec{x},\vec{x}_{0};z)
G^{(0)}(\vec{x},\vec{x}_{0};\lambda)d\vec{x}-
e^{i\theta}(z-\bar{\lambda})\int
G^{(0)}(\vec{x},\vec{x}_{0};z)
G^{(0)}(\vec{x},\vec{x}_{0};\bar{\lambda})d\vec{x}}.
\end{equation}
After substituting Eq.(\ref{2}) into Eq.(\ref{11}), and with  
a somewhat lengthy but straightforward calculation, we have 
\begin{equation}
\label{12}
T_{\theta}(z)=(v_{\theta}^{-1}-G(z))^{-1},
\end{equation}
where 
\begin{equation}
\label{13}
v_{\theta}^{-1}=
|\lambda|\frac{\sin(\frac{\theta}{2}+arg\lambda)}{\sin \frac{\theta}{2}}
\sum_{n=1}^{\infty}
\frac{\varphi_{n}(\vec{x}_{0})^{2}}{|E_{n}-\lambda|^{2}},
\end{equation}
\begin{equation}
\label{14}
G(z)=
\sum_{n=1}^{\infty}
\varphi_{n}(\vec{x}_{0})^{2}
(\frac{1}{z-E_{n}}+\frac{E_{n}}{|E_{n}-\lambda|^{2}}).
\end{equation}
It follows from Eq.(\ref{12}) that 
the perturbed eigenvalues are determined by 
\begin{equation}
\label{15}
G(z)=v_{\theta}^{-1}.
\end{equation}
The solutions of this equation, $z_n$, $(n=1,2,3,\cdots)$, correspond to 
the poles of the T-matrix. The corresponding eigenfunction 
is the residue of the Green's function of Eq.(\ref{10}) at the pole; 
\begin{equation}
\label{16}
\psi_n (\vec{x})=N_n G^{(0)}(\vec{x},\vec{x}_0;z_n)
\end{equation}
where the normalization factor is determined by  
\begin{equation}
\label{17}
N_n^{-2}=\sum_{k=1}^{\infty} 
\frac{\varphi_k (\vec{x}_0)^2}{(z_n-E_k)^2}.
\end{equation}

Although the T-matrix seemingly has two independent parameters, 
$\lambda$ and $\theta$, this is not the case.  
This follows from the fact that 
$|\lambda|$ is regarded as a scale of mass in case of 
$\lambda$ being pure imaginary. 
Indeed, taking $|\lambda|$ as a unit of energy, 
we can fix $\lambda=+i$, 
which makes all the relevant physical quantities to be dimensionless. 
Such treatment is justified by the assertion that 
a symmetric operator $T$ is self-adjoint 
if the condition $Ran(T-\lambda)=Ran(T-\bar{\lambda})=X$ 
is valid for {\em some fixed} $\lambda$ $(Im\lambda \neq 0)$. 
This means that if and only if $Ran(T\pm i)=X$,  
a symmetric operator $T$ is self-adjoint. 
Thus, without any loss of generality, we can make
the replacement of $|\lambda| \rightarrow 1$, 
$z \rightarrow |\lambda| z$, $E_n \rightarrow |\lambda| E_n$ 
and $\varphi_n \rightarrow  |\lambda| \varphi_n$ 
in Eqs.(\ref{13}), (\ref{14}) and (\ref{15}).
We then have   
\begin{equation}
\label{18}
\bar{G}(z)=\bar{v}_{\theta}^{-1}, 
\end{equation}
where 
\begin{equation}
\label{19}
\bar{v}_{\theta}^{-1}=
\frac{\sin \theta}{1-\cos \theta}
\sum_{n=1}^{\infty}
\frac{\varphi_{n}(\vec{x}_{0})^{2}}{E_{n}^2+1},
\end{equation}
\begin{equation}
\label{20}
\bar{G}(z)=
\sum_{n=1}^{\infty}
\varphi_{n}(\vec{x}_{0})^{2}
(\frac{1}{z-E_{n}}+\frac{E_{n}}{E_{n}^2+1}).
\end{equation}
We can regard $\bar{v}_{\theta}$ as a coupling constant 
of a pointlike scatterer. 
It ranges over all real numbers as $0 \leq \theta < 2\pi$. 

In a previous publication \cite{S94}, we have referred to the 
coupling constant 
$\bar{v}_{\theta}$ as the bare coupling constant, denoted by $v_B$.
However, this is somewhat confusing, because 
$\bar{v}_{\theta}$ should be considered to arise in a renormalization 
process for treating short-range singularities 
in a proper manner, which we often encounter in field theory. 
It might be more appropriate to call $\bar{v}_{\theta}$ the {\em renormalized 
coupling constant}. 

Because the average level density of 
two-dimensional billiards is independent of energy $z$, 
each term in the parenthesis of Eq.(\ref{20}) 
diverges when summed separately.
The divergence disappears when summed together. 
This means that the second term of Eq.(\ref{20}) 
plays an essential role to make the T-matrix well-defined.
It is also noteworthy that the energy dependence 
appears only in the first term of Eq.(\ref{20}). 
This ensures the orthogonality of the perturbed 
eigenfunctions $\{\psi_n (\vec{x})\}$, $(n=1,2,3,\cdots)$. 
Namely, for $m \neq n$, we have 
\begin{equation}
\label{21}
\int \psi_m (\vec{x}) \psi_n (\vec{x}) d\vec{x} =
N_m N_n \frac{\bar{G}(z_n)-\bar{G}(z_m)}{z_m - z_n} = 0. 
\end{equation}
The orthonormal relations of the unperturbed eigenfunctions 
$\{\varphi_n (\vec{x})\}$, $(n=1,2,3,\cdots)$, 
are used in the first equality, 
and the second equality results from of Eq.(\ref{18}). 
The authors of Ref.\cite{WS95} have discussed a rectangular billiard 
with a single pointlike scatterer and   
deduced an eigenvalue equation, 
Eq.(23) in Ref.\cite{WS95}, which is seemingly analogous 
to Eq.(\ref{18}) with Eqs.(\ref{19}) and (\ref{20}) 
in this paper. 
However, their approach allows the coupling strength
$\bar{v}_\theta$ to be varied depending on 
the energy, and violates the orthogonality, Eq.(\ref{21}).
\section{Quantum Spectrum of Pseudointegrable Billiards 
         with a Pointlike Scatterer}
Equipped with the formulation in the previous section, 
we examine the spectral properties 
of pseudointegrable billiards with a pointlike scatterer. 
A special emphasis is placed on the condition 
for the appearance of wave chaos. 

An important fact is that
each perturbed eigenvalue  
is isolated between two unperturbed ones. 
This follows from the fact that $\bar{G}(z)$ of Eq.(\ref{20}) is a
monotonously decreasing function of $z$ on the interval 
between any two successive unperturbed eigenvalues, $(E_n,E_{n+1})$,
covering the entire value between $(-\infty, \infty)$.
It is also clear that 
$\bar{G}(z)$ has a single inflection point on $(E_n,E_{n+1})$. 
We have shown in Ref.\cite{S94} that 
the disturbance by the pointlike scatterer 
is restricted to
the eigenstates with an eigenvalue around which 
$\bar{G}(z)$ has an inflection point (See Figs.2 and 3 
in Ref.\cite{S94}). 
This is understood also from the eigenfunction, Eq.(\ref{16}) 
with Eq.(\ref{2}); 
If a perturbed eigenvalue is close to an unperturbed one, 
the corresponding perturbed eigenfunction is not 
substantially different from the corresponding unperturbed 
one. 

In Ref.\cite{S94}, the condition for the inflection points has been 
deduced in a somewhat complicated manner, by making a 
truncation of the unperturbed basis according to the energy under 
consideration. 
Here, we refine the argument without introducing any truncation of the basis. 

Each inflection point of $\bar{G}(z)$, say $\tilde{z}_m$, 
$(m=1,2,3,\cdots)$, is expected to 
appear, in average, around the midpoint on $(E_m,E_{m+1})$;  
$\tilde{z}_m \simeq (E_m + E_{m+1})/2$.  
The contributions from $0<E_n <\tilde{z}_m$ 
and $\tilde{z}_m <E_n <2\tilde{z}_m \simeq E_{2m}$ 
on the summation of the first term in Eq.(\ref{20}) 
are canceled with high degree of accuracy. 
This allows us to estimate $\bar{G}(z)$ at 
the inflection point $\tilde{z}_m$ as follows; 
\begin{eqnarray}
\label{22}
\bar{G}(\tilde{z}_m) & \simeq &
\sum_{n=1}^{2m}
\varphi_{n}(\vec{x}_{0})^{2}\frac{E_{n}}{E_{n}^2+1} +
\sum_{n=2m+1}^{\infty}
\varphi_{n}(\vec{x}_{0})^{2}
(\frac{1}{\tilde{z}_m -E_{n}}+\frac{E_{n}}{E_{n}^2+1}) 
\nonumber \\
& \simeq & 
\alpha \{ \int_{0}^{E_{2m}} \frac{E}{E^2+1}dE + 
\int_{E_{2m}}^{\infty} (\frac{1}{\tilde{z}_m -E}+\frac{E}{E^2+1})dE \} 
\nonumber \\
& \simeq & \alpha \ln \tilde{z}_m, 
\end{eqnarray}
where we define 
$\alpha = \rho_{av} \langle \varphi_n (\vec{x}_0)^2 \rangle$ 
in terms of the average level density of the unperturbed system, 
$\rho_{av}$. 
The symbol $\langle \cdot \rangle$ means an average among various $n$. 
Because $\rho_{av} = MS/(2\pi)$ according to the Weyl's theorem 
and $\langle \varphi_n (\vec{x}_0)^2 \rangle \simeq 1/S$ 
for a generic choice of 
$\vec{x}_0$, we obtain 
\begin{equation}
\label{23}
\alpha = \rho_{av} \langle \varphi_n (\vec{x}_0)^2 \rangle \simeq 
\frac{M}{2\pi}. 
\end{equation}
It follows from Eqs.(\ref{18}) and (\ref{22}) that the effects 
of the pointlike scatterer 
on the quantum spectrum are observed in the eigenstates 
with an eigenvalue $z$ such that 
\begin{equation}
\label{24}
 \alpha \ln z \simeq \bar{v}_{\theta}^{-1}. 
\end{equation}
The condition of Eq.(\ref{24}) corresponds to Eq.(60) in Ref.\cite{S94}. 
It is also noteworthy that a $s$-wave phase shift of 
the two-dimensional scattering problem with a single pointlike scatterer 
shows a similar logarithmic dependence on energy $z$ \cite{AGHH88,J95}.  
The origin of such energy dependence 
is closely related to the energy scale $|\lambda|$ 
introduced by the self-adjoint extension of a 
symmetric operator; Any function which depends on $|\lambda|$ has 
an energy dependence so as to guarantee the argument to be dimensionless. 
Although the nature of a particle motion in billiards 
is independent of the energy in classical physics, 
the singularity of the interaction
induces an energy dependence of the observables 
after quantization. 
This can be considered as a typical example of 
quantum mechanical breaking of scale invariance, or 
{\em scale anomaly} \cite{CS95,J95}. 
 
Let us proceed to make an estimate of the width of 
a strip along the logarithmic curve of Eq.(\ref{24}) on which 
the disturbance of the scatterer is observable. 
Using the derivative of $\bar{G}(z)$ at an inflection point $\tilde{z}_m$, 
we can estimate the width, say $\Delta$, as follows; 
\begin{equation}
\label{25}
\Delta \ll |\bar{G}'(\tilde{z}_m)| \rho_{av}^{-1} \simeq
\langle \varphi_n (\vec{x}_0)^2 \rangle \sum_{n=1}^{\infty}
\frac{2}{\{(n-\frac{1}{2})\rho_{av}^{-1} \}^{2} }
\rho_{av}^{-1}. 
\end{equation}
We have implicitly assumed in Eq.(\ref{25}) that 
the unperturbed eigenvalues 
are distributed 
within a mean interval $\rho_{av}^{-1}$ 
in the whole energy region including negative energies. 
This assumption is quite satisfactory in this case, 
because the denominator of $\bar{G}'(z)$ is of the order of 
$(z-E_n)^2$, indicating that the summation in Eq.(\ref{25}) 
converges rapidly.  
With the aid of a relation
\begin{equation}
\label{26}
\sum_{n=1}^{\infty} \frac{1}{(2n-1)^2} =\frac{\pi^2}{8}, 
\end{equation}
we arrive at 
\begin{equation}
\label{27}
|\bar{G}'(\tilde{z}_m)| \rho_{av}^{-1} \simeq \pi^2 \alpha, 
\end{equation}
which means
\begin{equation}
\label{28}
\Delta \simeq \alpha. 
\end{equation}
The estimate of Eq.(\ref{28}) is justified  
from another perspective later in this paper. 

The above considerations lead us to the condition 
for the appearance of wave chaos.  
The effect of a pointlike scatterer with strength 
$\bar{v}_{\theta}$ appears mainly  
in the eigenstates with an eigenvalue $z$ satisfying     
\begin{equation} 
\label{29}
|\bar{v}_{\theta}^{-1}- \alpha \ln z| \lsimeq \Delta, 
\end{equation}
In other words, wave chaos occurs in the energy region 
which satisfies 
\begin{equation} 
\label{30}
\exp{\{\frac{1}{\alpha \bar{v}_{\theta}}-1\}} 
\lsimeq \ z \ \lsimeq 
\exp{\{\frac{1}{\alpha \bar{v}_{\theta}}+1\}}.  
\end{equation}
The condition of Eq.(\ref{29}) contains all the essential 
physics of wave chaos. From this, we can draw the following
conclusions.

(1) For any positive $\bar{v}_{\theta}$, wave chaos appears at the 
energy which satisfies Eq.(\ref{29}), while it is hardly seen 
in other energies.
If $\bar{v}_{\theta}$ is larger than $\Delta^{-1}$, 
wave chaos appears only around the ground state region. 
As the energy increases, 
it tends to disappear.  On the other hand, 
if $\bar{v}_{\theta}$ is substantially smaller than $\Delta^{-1}$, 
wave chaos appears in the higher energy region 
specified by Eq.(\ref{30}). 
In the limit of $\bar{v}_{\theta}  \rightarrow + 0$, the system 
restores the integrability by pushing up the chaotic region 
to the infinite energy. (In this limit, 
the eigenvalue of the single eigenstate with an eigenvalue smaller 
than $E_1$ diverges to $-\infty$.)

(2) Wave chaos does not appear at any energy if $\bar{v}_{\theta}$ is 
negative. In particular, the system converges to the unperturbed 
one as $\bar{v}_{\theta} \rightarrow - 0$.  

(3) For any $\bar{v}_{\theta}$, the system behaves as integrable 
in the high energy limit, which agrees with our intuition 
that a pointlike scatterer has no effects on a particle motion 
in the classical limit. 
In other words, the quantum billiards with a pointlike 
scatterer have the nature of {\em asymptotic freedom}, 
which has been first discovered in
the non-Abelian Gauge field theories.
\section{Formal Description of a Pointlike Scatterer 
         in terms of a Dirac's Delta Function} 
We have revealed  various features 
of the two-dimensional billiards with 
a single pointlike scatterer 
from a general perspective in the previous section. 
>From a practical point of view, however, 
we still need to identify   
the physical meaning of the coupling strength 
$\bar{v}_{\theta}$, which 
does not have a direct relation to physical observables. 
As a first step for this purpose, 
we show in this section that 
$\bar{v}_{\theta}$ can be related to the physical strength 
of a pointlike scatterer through a truncation of 
the unperturbed basis. 

Let us consider a {\em formal} Hamiltonian 
with a Dirac's delta function potential  
\begin{equation}
\label{31}
H = H_0 + v \delta (\vec{x}-\vec{x}_0)
\end{equation}
with a truncated unperturbed basis $\{ \varphi_n (\vec{x}) \}$, 
$(n=1,2,3,\cdots,N)$. 
We can regard $v$ as the physical coupling constant 
of the pointlike scatterer. 
Notice that the Hamiltonian as Eq.(\ref{31}) cannot be defined 
in the billiard problems with higher dimension than one because  
one cannot take the limit of $N \rightarrow \infty$. 
For a moment, however, we will go further along with fixed $N$ 
and return to this point later in this section. 

The characteristic equation of $H$ is given by 
\begin{equation}
\label{32}
det (z {\bf I} - {\bf H_{0}} - {\bf V} ) = 0,
\end{equation}
where ${\bf I}$ is a $N$-dimensional unit matrix. 
The matrix elements of ${\bf H_{0}}$ and ${\bf V}$ are 
given by 
\begin{equation}
\label{33}
({\bf H_{0}})_{mn} = E_{n} \delta_{mn}
\end{equation}
and 
\begin{equation}
\label{34}
({\bf V})_{mn} = v \varphi_{m}(\vec{x}_{0})\varphi_{n}(\vec{x}_{0})
\end{equation}
with $m, n=1,2,3,\cdots,N$, respectively. 
Using the linearity of a determinant 
and noticing that 
any two column vectors of ${\bf V}$ are linearly dependent, 
we obtain
\begin{equation}
\label{35}
det (z {\bf I} - {\bf H_{0}} - {\bf V} ) =  
\prod_{n=1}^{N} (z-E_{n}) - 
v \sum_{n=1}^{N} \{ \varphi_{n}^2 (\vec{x}_{0}) 
\mathop{\;\,{\prod}^{\prime}}_{m=1}^{N} (z-E_{m}) \}, 
\end{equation}
where $\mathop{\;\,{\prod}^{\prime}}$ signifies that 
the term of $m=n$ is removed from the product. 
By inserting Eq.(\ref{35}) into Eq.(\ref{32}), 
the characteristic equation can be rewritten as  
\begin{equation}
\label{36}
\tilde{G}(z)=v^{-1},
\end{equation}
where 
\begin{equation}
\label{37}
\tilde{G}(z)= \sum_{n=1}^{N} 
\frac{\varphi_{n}(\vec{x}_0)^2}{z-E_n}.
\end{equation}
For any $v$, Eq.(\ref{36}) has $N$ non-degenerate solutions 
$z_n$, $(n=1,2,3,\cdots,N)$, giving a complete set of eigenvalues 
of the Hamiltonian Eq.(\ref{31}) with the truncated basis. 
The corresponding eigenfunction is given by 
\begin{equation}
\label{38}
\psi_n (\vec{x})=N_n \sum_{k=1}^{N} 
\frac{\varphi_{k}(\vec{x}) \varphi_{k}(\vec{x}_0)}
     {z_n-E_k}
\end{equation}
with the normalization factor determined by  
\begin{equation}
\label{39}
N_n^{-2}=\sum_{k=1}^{N} 
\frac{\varphi_k (\vec{x}_0)^2}{(z_n-E_k)^2}. 
\end{equation}
The eigenfunction, Eq.(\ref{38}) has a form analogous to 
Eq.(\ref{16}). 
For large $N$, the appropriate boundary condition around the scatterer  
is taken into account in a matrix diagonalization 
with a Dirac's delta function potential. 

In an analysis similar to the previous section, 
we recognize that  
the inflection points of $\tilde{G}(z)$ are located 
around $\tilde{G}(z) \simeq 0$ at energy $z \simeq E_N /2$, 
implying the appearance of wave chaos 
under the condition of $v^{-1} \simeq 0$. 
We also expect that it disappears if $|v^{-1}| \gsimeq \Delta$.  
In case of $|v^{-1}| \simeq \Delta$, 
the average ratio between the diagonal matrix element of 
the delta function potential and the mean level spacing is 
estimated as 
\begin{equation}
\label{40}
\left| \frac{\langle ({\bf V})_{nn} \rangle}{\rho_{av}^{-1}} \right| = 
\frac{\alpha}{\Delta} 
\simeq 1. 
\end{equation}
This explains the reason why the logarithmic curve of Eq.(\ref{24}) 
has a strip with width $\Delta$; 
In case of $|v^{-1}| \gsimeq \Delta$, the physical strength $v$ is too weak 
to cause the mixing among a large number of unperturbed eigenstates.  
In this sense, we can regard $|v^{-1}|$ as a measure of distance 
to the wave-chaotic curve. 
We stress that $|v^{-1}| \simeq 0$ corresponds to the condition for 
the appearance of wave chaos in the energy region around $z \simeq E_N /2$. 
Both at the lower and higher energies, 
the value of $\tilde{G}(z)$ at the inflection points 
is substantially different from zero, 
since the contributions on $\tilde{G}(z)$ 
from the terms with $E_n <z$ and those with $E_n >z$ 
do not cancel. 
This fact gives rise to an interesting physical result 
in the following section. 

Eq.(\ref{37}) allows us to relate the physical strength $v$ to 
the formal strength $\bar{v}_{\theta}$ introduced in Sec.II. 
In a similar manner as in Eq.(\ref{22}), we get
for $z < E_N$,     
\begin{eqnarray}
\label{41}
\bar{G}(z) & = & \tilde{G}(z)+ 
\sum_{n=1}^{N}
\varphi_{n}(\vec{x}_{0})^{2}\frac{E_{n}}{E_{n}^2+1} +
\sum_{n=N+1}^{\infty}
\varphi_{n}(\vec{x}_{0})^{2}(\frac{1}{z-E_{n}}+\frac{E_{n}}{E_{n}^2+1}) 
\nonumber \\
& \simeq & 
\tilde{G}(z) + \alpha \{ \int_{0}^{E_N} \frac{E}{E^2+1}dE + 
          \int_{E_N}^{\infty} (\frac{1}{z-E}+\frac{E}{E^2+1})dE \} 
\nonumber \\
& \simeq & \tilde{G}(z) + \alpha \ln (E_N - z). 
\end{eqnarray}
>From Eqs.(\ref{18}), (\ref{36}) and (\ref{41}), 
we obtain the relation between both the coupling constants;  
\begin{equation}
\label{42}
\bar{v}_{\theta}^{-1} \simeq v^{-1}+\alpha \ln (E_N - z). 
\end{equation}

We note that the summation in Eq.(\ref{37}) diverges 
as $N$ increases for any $z \neq E_n$. 
This follows from the fact that $\alpha$ defined in Eq.(\ref{23}) 
is independent of energy $z$. 
This underscores the fact that the Dirac's delta function potential 
with a full unperturbed basis 
cannot be defined in two-dimensional billiard problems, 
contrary to in one dimension. 
In the latter case, since $\alpha \propto \rho_{av}(z) \propto z^{1/2}$,  
the RHS of Eq.(\ref{37}) 
converges. 
In this case, one can relate $v$ with $\bar{v}_{\theta}$ through 
\begin{equation}
\label{43}
v^{-1} = \bar{v}_{\theta}^{-1} -
\sum_{n=1}^{\infty}
\varphi_{n}(\vec{x}_{0})^{2}\frac{E_{n}}{E_{n}^2+1}, 
\end{equation}
to make the matrix diagonalization with strength $v$ identical to 
the formulation with $\bar{v}_{\theta}$ discussed in Sec.II. 
Thus it is possible to construct quantum mechanics of 
one-dimensional billiards with a pointlike scatterer
directly in terms of a Dirac's delta function 
without relying on a somewhat complicated mathematical 
framework based on functional analysis. 
We finally emphasize that 
although the Hamiltonian 
with a Dirac's delta function potential, Eq.(\ref{31}), 
loses its meaning in the limit of $N \rightarrow \infty$ 
in two-dimensional billiards, 
Eq.(\ref{42}) is still valid for any $v$, $\bar{v}_{\theta}$ 
if $z < E_N$ with an arbitrarily fixed $N$. 
\section{Quantum Spectrum of Billiards with a Small but 
         Finite-size Scatterer} 
In this section, we discuss the quantum spectra of 
the billiards with a small but finite-size scatterer. 
We are after the wave-chaotic 
spectra in the systems with a realistic 
finite-size scatterer inside. 
For this purpose, we attempt to describe 
the finite-size scatterer 
in terms of a Dirac's delta function potential 
with a suitably truncated basis. 
This also enables us to recognize the physical meaning of 
the formal coupling constant $\bar{v}_{\theta}$. 
Relying on the arguments in Secs.III and IV, 
we clarify the condition for the appearance of wave-chaotic 
spectra for billiards with a small scatterer. 

Suppose that a finite-size scatterer of area $\Omega$ is placed at   
$\vec{x}=\vec{x}_0$ inside an integrable billiard of area $S$. 
We describe the interaction in terms of a potential with a constant 
strength on the domain of the scatterer;  
\begin{equation}
\label{44}
V(\vec{x}) = 
\left \{
\begin{array}{ll}
V, & \vec{x} \in \Omega, \\
0, & \vec{x} \in  S-\Omega, 
\end{array}
\right. 
\end{equation}
where we denote the domains of the scatterer and the outer billiard by 
the same symbols as the areas. 
The matrix elements of $V(\vec{x})$ are given by 
\begin{equation}
\label{45}
({\bf V})_{mn} = V \int_{\Omega} 
\varphi_{m}(\vec{x})\varphi_{n}(\vec{x}) d\vec{x}, \ \ \
(m, n=1,2,3,\cdots). 
\end{equation}

If the area of the scatterer is far smaller than that 
of the outer billiard, 
the scatterer is expected to behave as pointlike at low energy 
because waves with long-wavelength cannot see 
the shape of the scatterer. 
More precisely, the matrix element $({\bf V})_{mn}$ 
with $E_m, E_n \lsimeq (2M \Omega)^{-1}$ 
is suitably approximated by $V\Omega \varphi_m (\vec{x}_0) 
\varphi_n (\vec{x}_0)$. 
Furthermore, the off-diagonal matrix elements with 
$E_m \ll (2M \Omega)^{-1} \ll E_n$ are 
small, because of the mismatch in  the wave numbers. 
These indicate that the energy spectrum of the billiards 
with a finite-size scatterer of area $\Omega$ 
can be reproduced by the Hamiltonian Eq.(\ref{31}) 
in terms of a Dirac's delta function potential with strength
\begin{equation}
\label{46}
v(\Omega) = V\Omega 
\end{equation}
along with a restricted basis up to 
\begin{equation}
\label{47}
E_{N(\Omega)} \simeq (2M\Omega)^{-1}. 
\end{equation}
It is crucial that the zero-range approximation 
is justified as long as $z \ll E_{N(\Omega)}$,  
even if $\left| V \right|$ is sufficiently large. 
The mixing among unperturbed eigenstates  
is limited to a small number of states
even if $v$ (or $V$) is sufficiently large. 
Note also that the matrix {\bf V} has a band-like structure, 
the width of which is independent of $V$, but is 
essentially determined by $\Omega$. 

We can now describe the low-energy spectra 
of the billiards with a finite-size scatterer inside 
within the framework in Sec.II.  
It is realized from Eq.(\ref{42}) that, 
as long as $z \ll E_{N(\Omega)}$, 
the eigenvalues in case of finite-size scatterer 
can be calculated by Eq.(\ref{18}) with 
\begin{eqnarray}
\label{48}
\bar{v}_{\theta}^{-1} & \simeq & 
v(\Omega)^{-1}+\alpha \ln (E_{N(\Omega)}-z) \nonumber \\
& \simeq &v(\Omega)^{-1}+\alpha \ln E_{N(\Omega)}.  
\end{eqnarray}
The meaning of the renormalized coupling constant  
$\bar{v}_{\theta}$ becomes clear; A quantum billiard with 
a pointlike scatterer with strength $\bar{v}_{\theta}$ can be 
obtained by taking the limit $\Omega \rightarrow 0$ along with 
the limit
\begin{equation}
\label{49}
v(\Omega) = \frac{1}{\bar{v}_{\theta}^{-1}+\alpha \ln (2M\Omega)}
\rightarrow -0, 
\end{equation}
or equivalently, with the limit
\begin{equation}
\label{50}
V = \frac{1}{\bar{v}_{\theta}^{-1}\Omega+\alpha \Omega\ln (2M\Omega)} 
\rightarrow - \infty.  
\end{equation}
Any two sets of $(V,\Omega)$, 
say $(V_1,\Omega_1)$ and $(V_2,\Omega_2)$, which satisfy  
Eq.(\ref{50}) with fixed $\bar{v}_{\theta}$, describe the same 
low-energy dynamics as long as $z \ll E_{N(\Omega)}$, 
where $\Omega=\max \{\Omega_1,\Omega_2 \}$. 
The logarithmic term in the denominator of Eq.(\ref{50}) expresses 
the reason why we cannot define a pointlike scatterer in terms of 
a Dirac's delta function in two dimension; 
The limit of $\Omega \rightarrow 0$ 
along with keeping $V\Omega$ constant 
induces a potential which is too strong to define a quantum 
mechanical Hamiltonian for a pointlike scatterer. 
It is noteworthy that 
in the small-size limit, the strength of the scatterer
is always negative in the sense that 
$V \rightarrow -\infty$ as $\Omega \rightarrow 0$. 
This is consistent with the fact that  
a single eigenstate with an eigenvalue smaller than $E_1$ 
exists for any $\bar{v}_{\theta}$.
We emphasize that this is not due to a specific formulation 
discussed here, but is required from the self-adjointness 
of the Hamiltonian. 
It should be noted, however, that 
as long as the obstacle has a substantial size, 
we can deduce all the low-energy physics 
from the present formulation by identifying $v(\Omega)$ 
with $\bar{v}_{\theta}$ through Eq.(\ref{48}), 
even in the case of positive $V$.
  
Inserting Eq.(\ref{48}) to Eq.(\ref{29}), we obtain 
the condition for the appearance of the wave-chaotic 
energy spectrum; 
\begin{equation} 
\label{51}
\left| v(\Omega)^{-1} - \alpha \ln \frac{z}{E_{N(\Omega)}-z} \right| 
\lsimeq \Delta \simeq \alpha. 
\end{equation}
Keeping Eq.(\ref{23}) in mind, 
we can rewrite Eq.(\ref{51}) as 
\begin{equation} 
\label{52}
\left| \left( 
\frac{\langle ({\bf V})_{nn} \rangle}{\rho_{av}^{-1}} \right) ^{-1} 
- \ln \frac{z}{E_{N(\Omega)}-z} \right| \lsimeq 1. 
\end{equation}
Because the logarithmic function in Eq.(\ref{51}) is negative 
for $z < E_{N(\Omega)}/2$, 
we obtain an interesting spectral property for lowest eigenstates; 
The wave chaos is most visible when the potential 
is attractive.  
As the energy increases up to $z \simeq E_{N(\Omega)}/2$, 
the wave-chaotic region is expected to shift 
to $|v(\Omega)^{-1}| \rightarrow 0$ according to Eq.(\ref{51}). 
However, the zero-range approximation tends to lose its validity 
at high energy, particularly in attractive cases as shown later 
in the numerical calculations. 

The previous numerical observations of Sinai's billiard
give a supplementary evidence for the above finding.   
In Ref.\cite{B81}, Berry has exhibited the energy-level diagram 
of Sinai's billiard as a function of the radius of 
a circular obstacle in the billiard. It reveals that 
the integrability is restored in the low energy region 
as the radius decreases. 
This can be easily understood in the present scope. 
When the potential is repulsive, the wave chaos never appears  
even at low energy 
even if the repulsion is sufficiently strong. 
In spite of the infinite height of the potential
in Sinai's billiard, 
the limit of the radius being zero can be taken without any difficulty 
because only the `desymmetrized states' are considered in Ref.\cite{B81}. 

In order to confirm the validity of Eq.(\ref{51}), 
we examine the quantum spectra of a rectangular billiard with 
a small rectangular scatterer inside, which is analogous to 
the system discussed in Ref.\cite{CC89}. 
The unperturbed eigenvalues and the corresponding normalized 
eigenfunctions of the empty billiard 
are given by 
\begin{equation}
\label{53}
E_{mn} = \frac{1}{2M}
\{(\frac{m\pi}{l_x})^2+
(\frac{n\pi}{l_y})^2 \},
\end{equation}
and
\begin{equation}
\label{54}
\varphi_{mn}(x,y) =
\sqrt{\frac{4}{l_x l_y}}
\sin \frac{m\pi x}{l_x}
\sin \frac{n\pi y}{l_y}, 
\end{equation}
with $m,n=1,2,3,\cdots$, respectively. 
The side lengths of the (outer) rectangle are denoted by 
$l_x$ and $l_y$. 
Suppose that a rectangular scatterer is placed 
inside the outer rectangle 
such that the sides of the inner and outer rectangles 
are parallel to each other. 
We denote
the side lengths and 
the position of the center of the scatterer 
by $\delta l_x$, $\delta l_y$ and 
$(x_0,y_0)$, respectively. 
The matrix elements of the potential of Eq.(\ref{44}) 
are given by 
\begin{equation}
\label{55}
({\bf V})_{m_1,n_1;m_2,n_2} = V 
u_{m_1,m_2} (l_x,\delta l_x,x_0)
u_{n_1,n_2} (l_y,\delta l_y,y_0),
\end{equation}
where  
\begin{equation}
\label{56}
u_{m,n} (l,\delta l,x_0) =
\left \{
\begin{array}{ll}
\frac{2}{\pi} 
\{ \frac{1}{m-n} \cos \frac{(m-n)\pi x_0}{l} 
\sin \frac{(m-n)\pi \delta l}{2l} & \\
\ \ \ \  - \frac{1}{m+n} \cos \frac{(m+n)\pi x_0}{l} 
\sin \frac{(m+n)\pi \delta l}{2l} \}, & m \neq n, \\
\frac{\delta l}{l}- 
\frac{1}{n\pi} \cos \frac{2n\pi x_0}{l} 
\sin \frac{n\pi \delta l}{l}, & m=n. 
\end{array}
\right. 
\end{equation}
In the actual numerical calculations, 
we set $M=2\pi$, $l_x = \pi/3 = 1.04719$ and 
$l_y = 3/\pi = 0.954929$, 
which leads to $\rho_{av}=1$ and $\alpha \simeq 1$. 
We also assume $\delta l_x = 3.53830 \times 10^{-2}$, 
$\delta l_y = 3.14023 \times 10^{-2}$ and 
$\vec{x}_0=(x_0,y_0)=(0.622482,0.275835)$. 
In this choice, the area of the scatterer, $\Omega$, 
is $1/900$ of the outer rectangle. 

Fig.1(a) shows several lowest eigenvalues of this system 
for various values of the inverse physical strength. 
The circles indicate the exact eigenvalues obtained by   
diagonalizing the Hamiltonian matrix along with 
Eq.(\ref{55}). 
The strength of the scatterer, $V$, is 
determined by Eq.(\ref{46}). The value of $v^{-1}$ ranges 
from $-10$ to $10$ at intervals 
of $1$. 
The solid curves are the eigenvalues obtained by 
the approximation in terms of a Dirac's delta function potential. 
In the latter case, 
we truncate the unperturbed basis at
$E_{N(\delta l_x,\delta l_y)}=
\{ (\pi/2 \delta l_x)^2 + (\pi/2 \delta l_y)^2 \}/(2M) = 355.949$, 
irrespective to the value of $v$. 
This corresponds to making a truncation at the momenta
$k_x \simeq \pi/(2\delta l_x)$ and 
$k_y \simeq \pi/(2\delta l_y)$. 
>From Fig.1(a), 
we observe that the approximation of a zero-range potential 
with a suitably truncated basis is quite satisfactory,  
even if $v$ is quite large. 
Since we have
$v \simeq \langle ({\bf V})_{m,n;m,n} \rangle / \rho_{av}^{-1}$ 
with $\alpha \simeq 1$, 
$v^{-1}=0$ ($v=1000$) indeed 
corresponds to a strong repulsion in the sense that 
the matrix elements of ${\bf V}$ are far larger than 
$E_{N(\delta l_x,\delta l_y)}$.  
Exceptions are the cases of a strong attractive potential. 
In this case, there is a considerable discrepancy 
between the exact and approximated eigenvalues. 
Fig.1(b) shows the eigenvalues for large $|v|$.  
When the scatterer is attractive, 
there exist, in general, several eigenstates 
with negative eigenvalues; 
For example, three eigenstates with eigenvalues 
$-1098$, $-261$ and $-151$ for $v^{-1}=-0.5$, 
and four eigenstates with eigenvalues 
$-2182$, $-1131$, $-944$ and $-12$ for $v^{-1}=-0.3$. 
(These values, particularly for shallow eigenstates 
are not fully convergent despite 
a huge matrix diagonalization with 12000 unperturbed states.) 
The eigenvalue of negative eigenstates is sensitive to 
the value of $v$. 
This causes a rapid change of the lowest positive eigenvalues 
as $v$ varies. 

As shown in Fig.1(a), the inflection points
of the solid curves 
in the low energy region 
appear in a weakly attractive region. 
The prediction of Eq.(\ref{51}) shown by a strip  
between two broken lines in Fig.1(a) reproduces the position 
of the inflection points. 
In Fig.2, we show the contour plot of the exact eigenfunction 
for several eigenstates. It can be seen that 
the mixture of the unperturbed eigenfunctions 
indeed occurs around the inflections points, 
while the wavefunction does not differ substantially 
from one of the unperturbed eigenfunctions 
in the region far from the inflection points. 
We stress that, even if $|v|$ is sufficiently large, 
the effects of high momentum components are 
visible mainly
in the vicinity of the scatterer. 
The global behavior of eigenfunctions gradually changes 
along the solid curves as $v^{-1}$ varies. 
It is expected that, 
as the scatterer shrinks ($E_{N(\delta l_x,\delta l_y)}$ 
becomes large), the wave-chaotic region spreads over to higher 
energy. Moreover, the effective strength of the scatterer 
becomes weaker with smaller value for $|v|$ 
for a fixed energy in 
$z \ll E_{N(\delta l_x,\delta l_y)}$, 
as indicated in Eq.(\ref{51}). 

Finally, we examine the validity of the zero-range approximation 
at higher energy. 
Figs.3-5 show the eigenvalues in three cases; 
$z \simeq  0.10 E_{N(\delta l_x,\delta l_y)}$ in Fig.3, 
$z \simeq  0.30 E_{N(\delta l_x,\delta l_y)}$ in Fig.4 and 
$z \simeq  0.55 E_{N(\delta l_x,\delta l_y)}$ in Fig.5, 
respectively. 
It is observed that the accuracy of the zero-range approximation 
depends on the sign of the potential. 
For attractive cases, the zero-range approximation fails 
at somewhat low energy, as pointed out before. 
There is a considerable discrepancy even for $v \simeq -1$ 
at $z \simeq  0.10 E_{N(\delta l_x,\delta l_y)}$.  
On the other hand,  for repulsive cases,
the approximated eigenvalues are in fairly good 
agreement with the exact eigenvalues up to 
$z \simeq  0.30 E_{N(\delta l_x,\delta l_y)}$. 
As the energy increases further, the zero-range approximation 
starts to fail
except for in case of a very weak potential. 
It is easily recognized that 
the spectral properties at high energy is subject to 
the geometry of the scatterer,  
since a quantum 
particle moving in the billiard tends to see the shape 
of the scatterer as the energy increases. 
\section{Conclusion}
We have examined the quantum spectra
in two-dimensional billiards 
with a small-size scatterer inside. 
First, we have clarified the spectral  
properties of pseudointegrable billiards with 
a pointlike scatterer from a general point of view. 
The strength of a pointlike scatterer is 
specified by a renormalized coupling constant 
$\bar{v}_{\theta}$, 
formally defined within the formulation based on 
functional analysis. 
Although the coupling constant does not have 
a direct relation to physical observables, 
it can be related to the physical coupling constant 
$v$ defined as a strength 
of a Dirac's delta function potential together with a truncated basis. 
In two-dimensional billiard problems, 
the zero-range interaction cannot be rigorously described 
in terms of a Dirac's delta function. 
Nevertheless, $v$ can be related to $\bar{v}_{\theta}$ 
as long as the unperturbed basis is truncated at
a finite number. From a physical perspective, 
the truncation of the basis is a consequence of  
finiteness of the size of the scatterer, 
as the uncertainty principle implies. 
Applying the findings in pseudointegrable billiards  
to the cases of a small but finite-size scatterer, 
we have shown that the signature of wave chaos 
in lowest eigenstates 
are observed most prominently 
when the scatterer is weakly attractive. 
The numerical experiments of a rectangular 
billiard with a small rectangular scatterer inside 
corroborate these arguments.

\acknowledgements

One of the authors (T.S.) acknowledges the support 
of the Grant-in-Aid for Encouragement of Young Scientists 
(No.07740316) by the Ministry of Education, Science, 
Sports and Culture. 
Numerical calculations have been performed on 
HITAC S-3800 supercomputer at the Computer Centre, 
the University of Tokyo. 
%
%\newpage
%

%
%

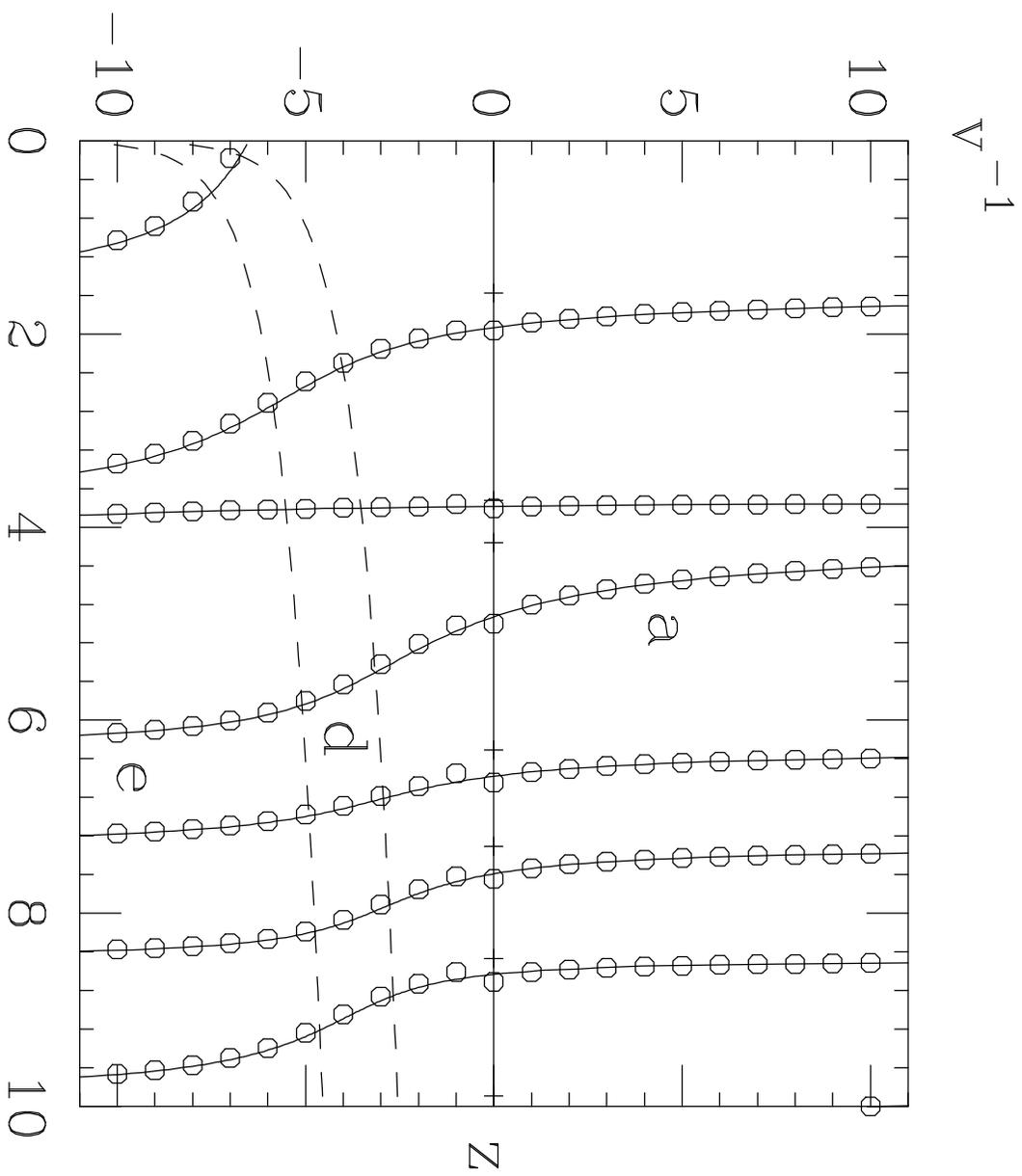
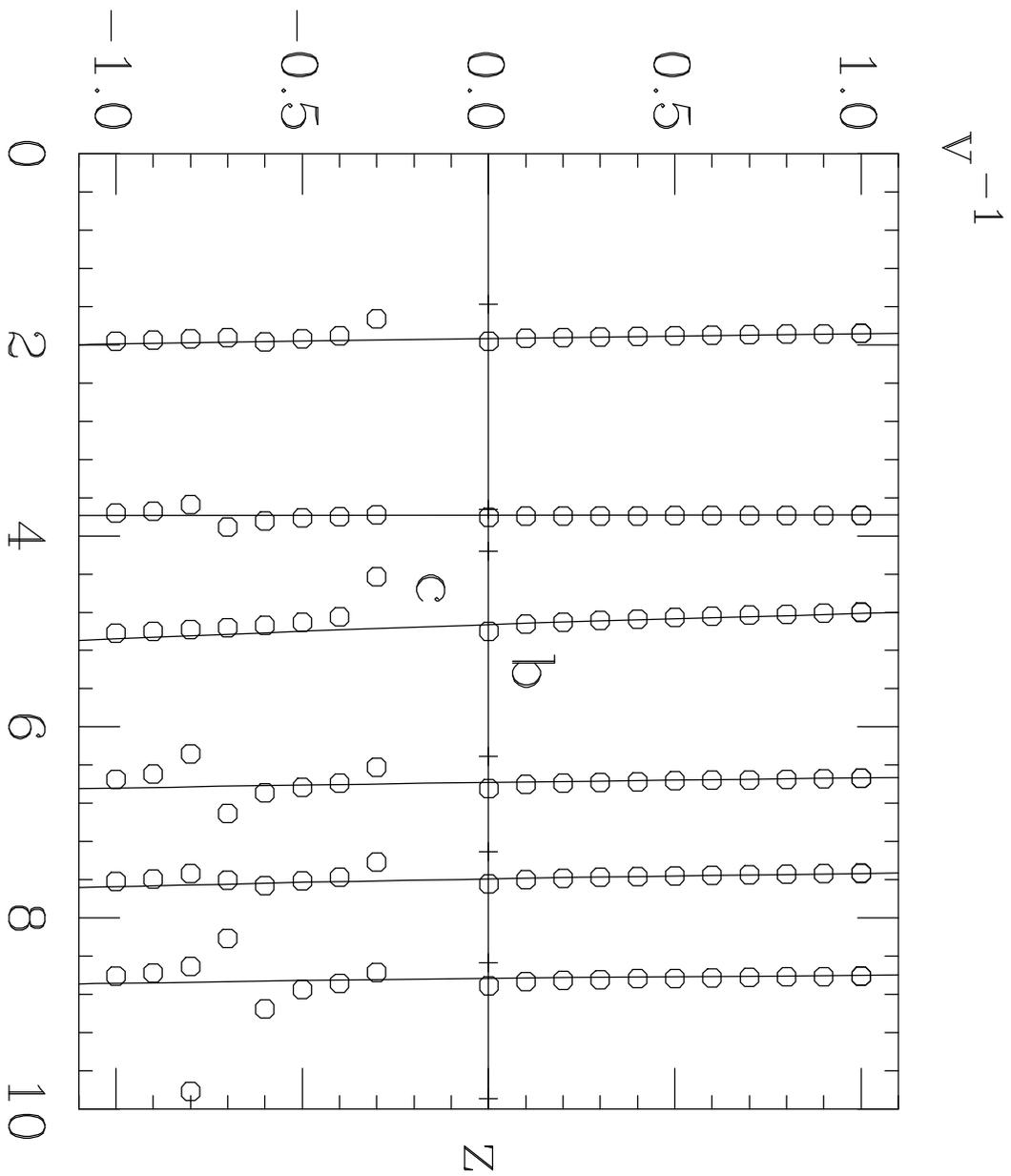
\begin{figure}
\caption{
Several lowest eigenvalues are shown for 
various values of the inverse physical strength 
of the scatterer, $v^{-1}$. 
The circles indicate the exact eigenvalues obtained by   
diagonalizing the Hamiltonian matrix along with 
Eq.(55). In this case, $V=v / \Omega$ with $\Omega=
1/900$. 
While the value of $v^{-1}$ ranges from $-10$ to $10$ 
at intervals of $1$ in (a), it ranges from 
$-1$ to $-0.3$ and from $0$ to $1$ at 
intervals of $0.1$ in (b). The case of $v^{-1}=0$ is 
calculated by taking $v=1000$. 
The solid curves are the approximated eigenvalues 
in terms of a Dirac's delta function potential 
with strength $v$, where 
the unperturbed eigenfunctions are truncated at   
$E_{N(\delta l_x,\delta l_y)}=
\{ (\pi/2 \delta l_x)^2 + (\pi/2 \delta l_y)^2 \}/(2M)$, 
irrespective to the value of $v$. 
Unperturbed energies are indicated by vertical lines 
on the $z$ axis. 
The strip between two broken lines in (a) is a prediction 
of Eq.(51) with $\alpha=1$, on which 
the signature of wave chaos appears. 
}
\end{figure}
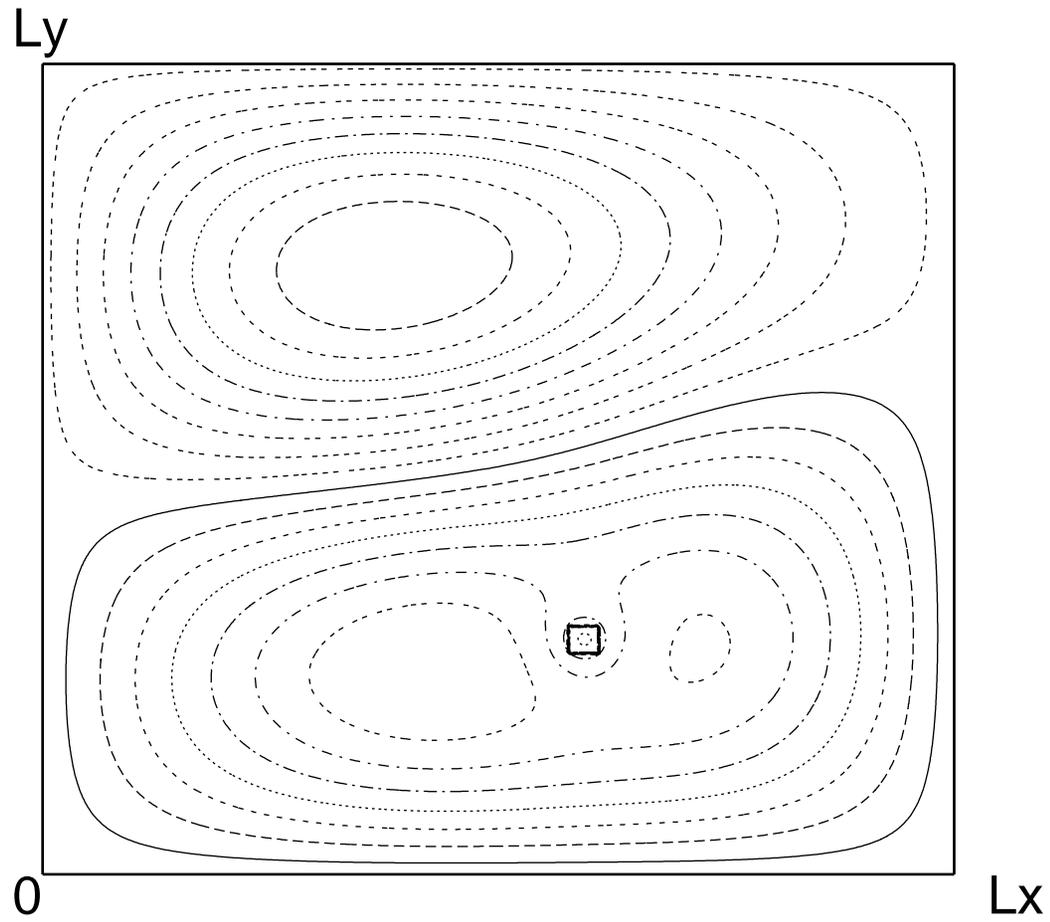
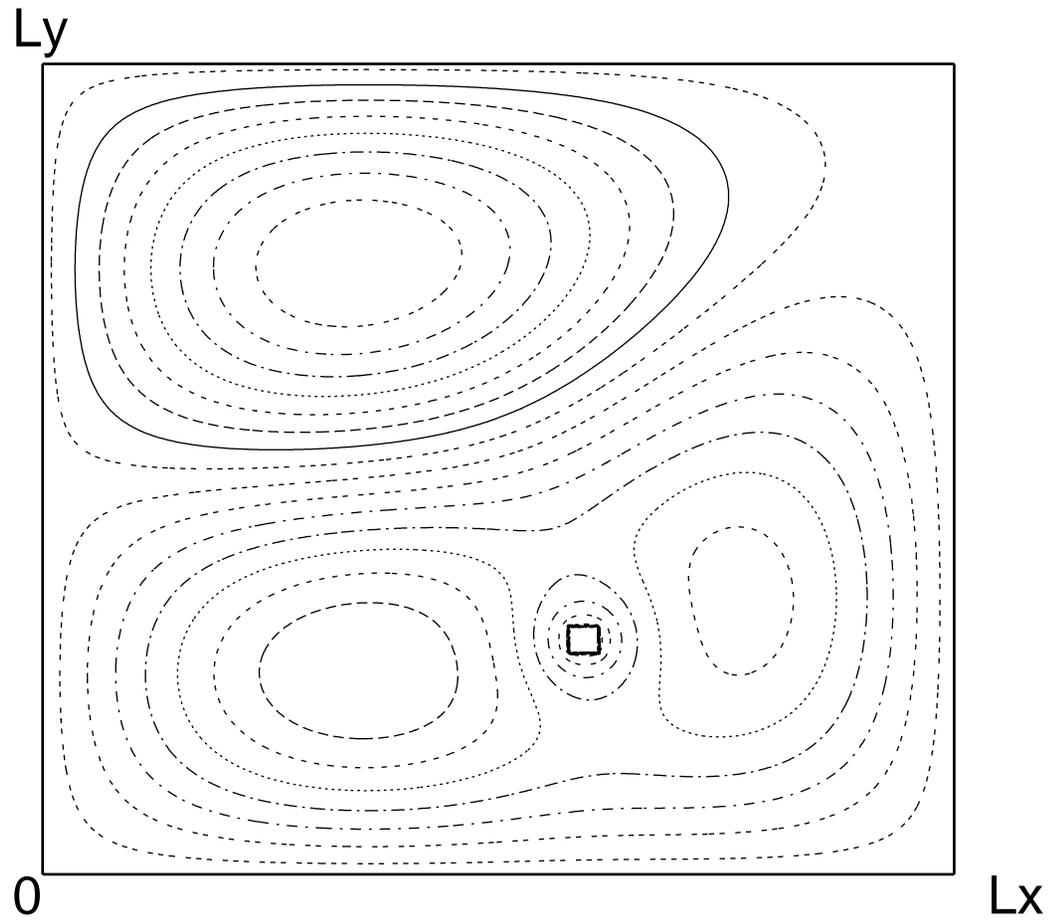
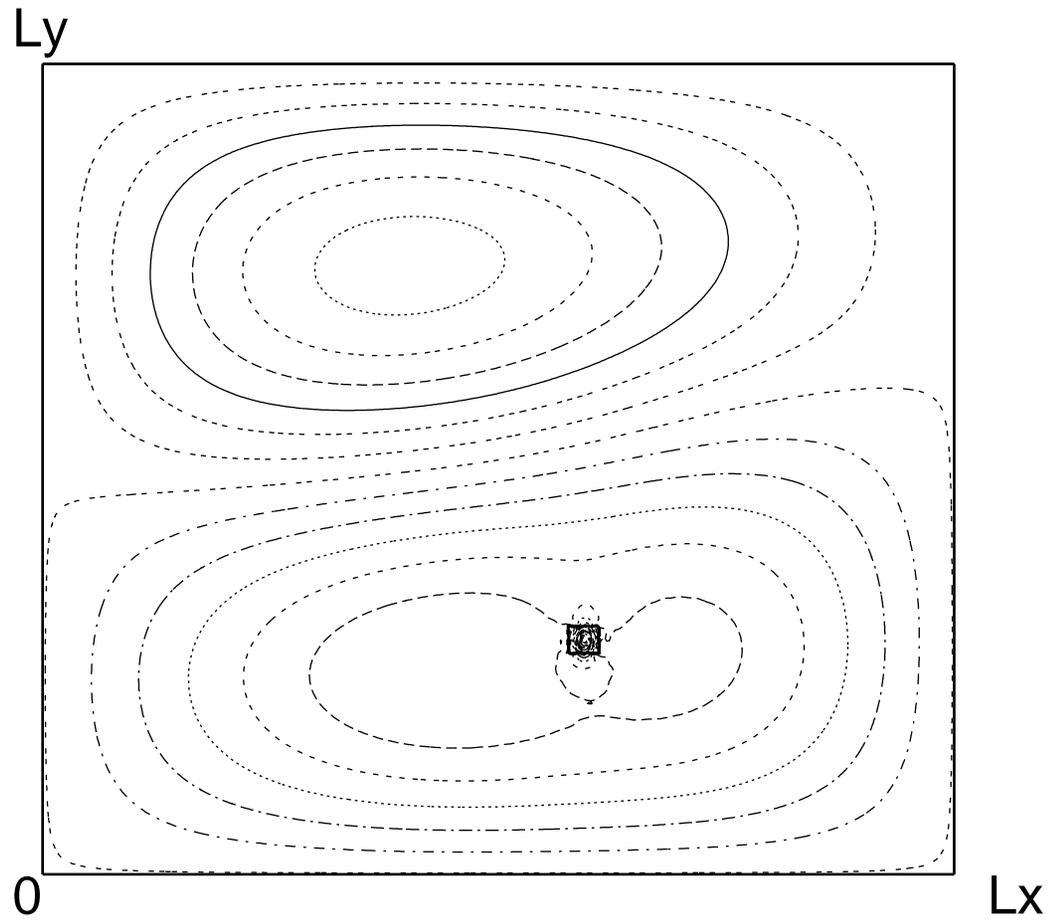
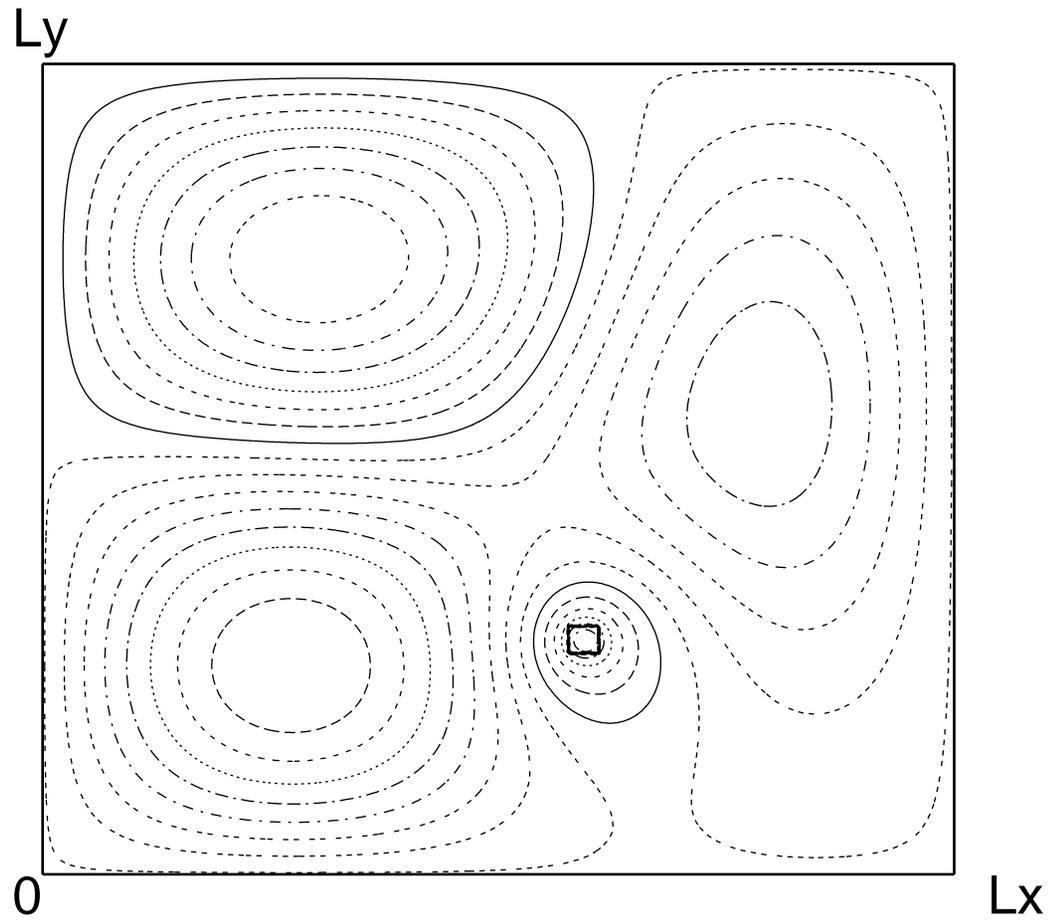
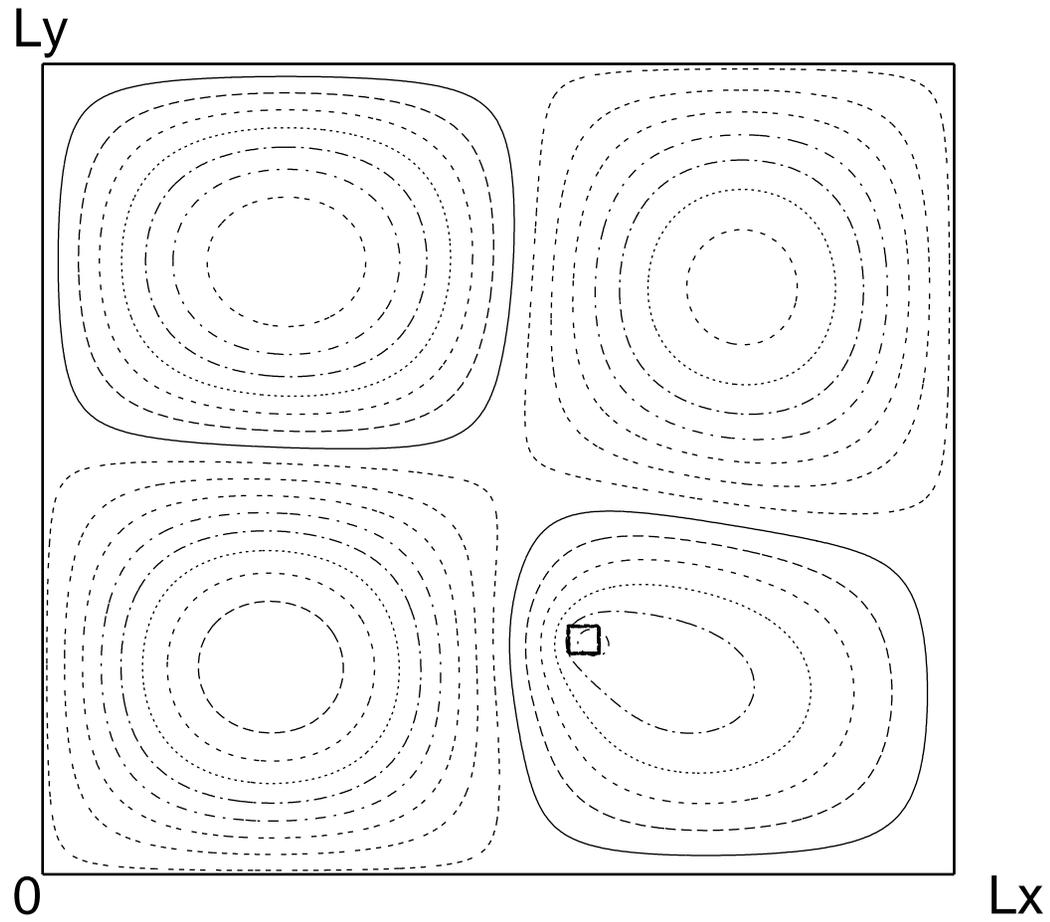
\begin{figure}
\caption{
The contour plot of the wavefunction for the eigenstates 
indicated by $a$-$e$ in Fig.1 is shown in (a)-(e), 
respectively. The value of $(v^{-1},z)$ for $a$-$e$ is 
$(5.0,4.54)$, $(0.1,4.93)$, $(-0.3,4.43)$, 
$(-4.0,5.63)$ and $(-10.0,6.13)$, respectively. 
The location of the scatterer is indicated by a rectangle 
in the billiard. 
}
\end{figure}
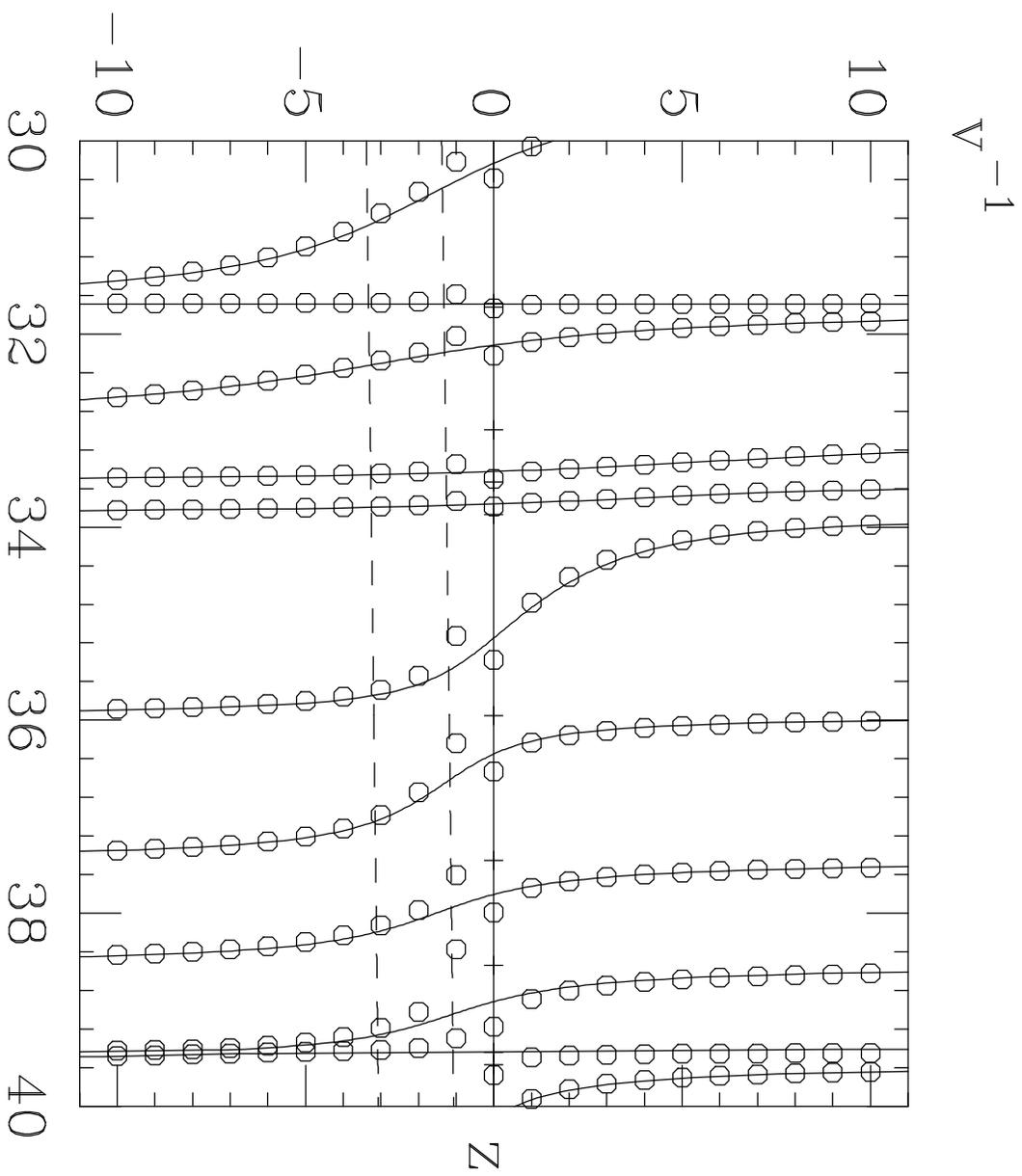
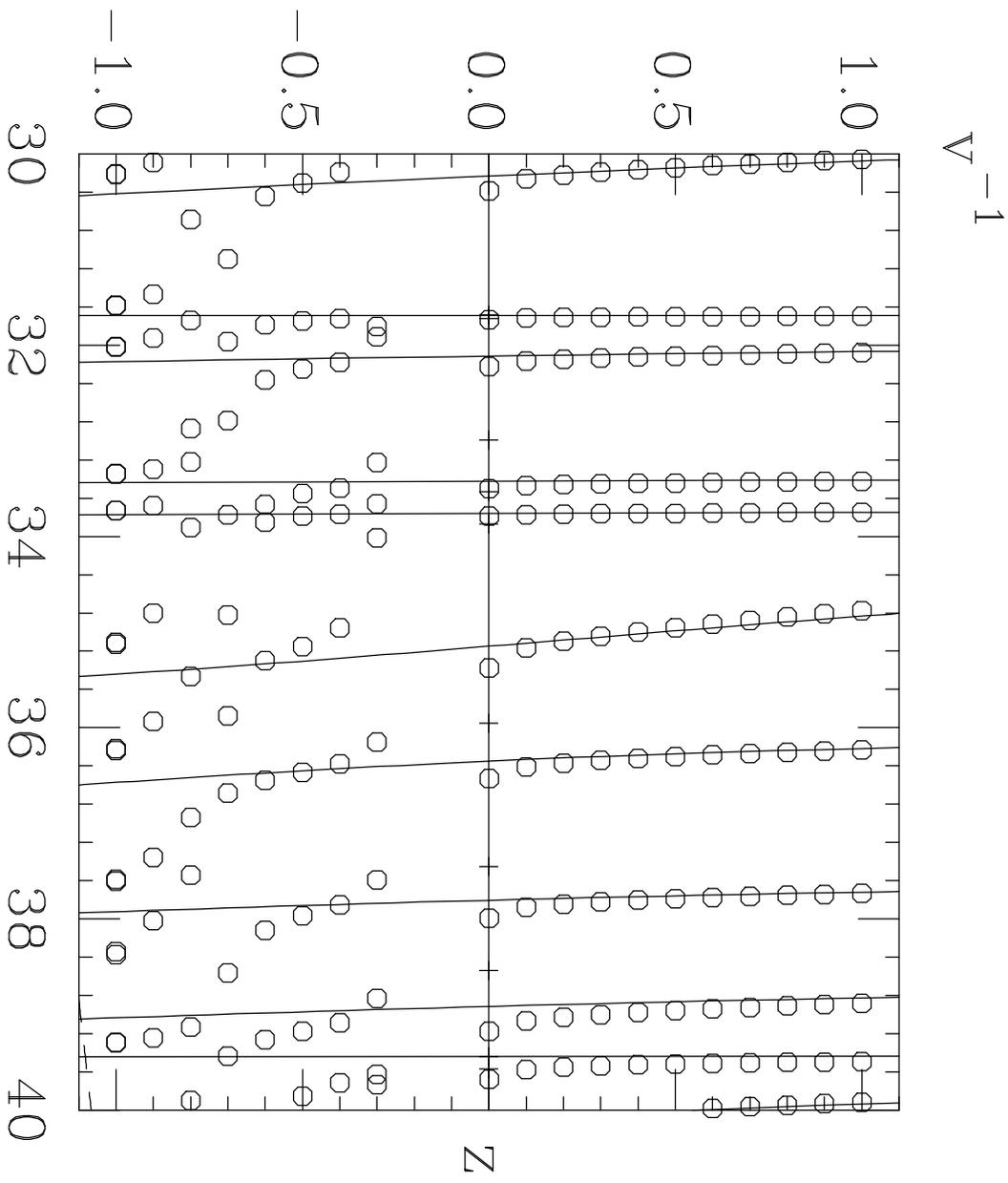
\begin{figure}
\caption{
The eigenvalues around  
$z \simeq  0.10 E_{N(\delta l_x,\delta l_y)}$.  
The indications are the same as in Fig.1.  
}
\end{figure}
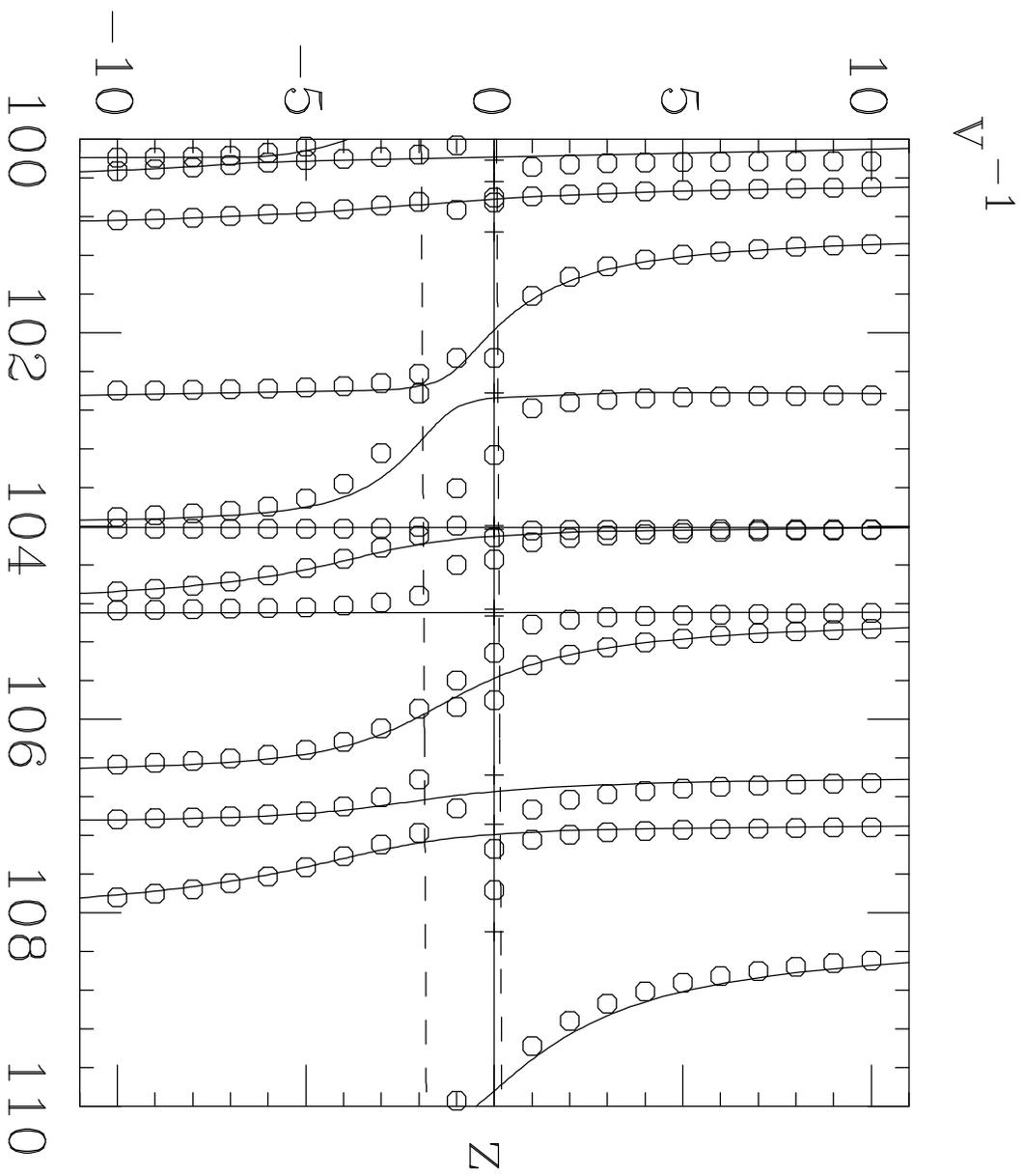
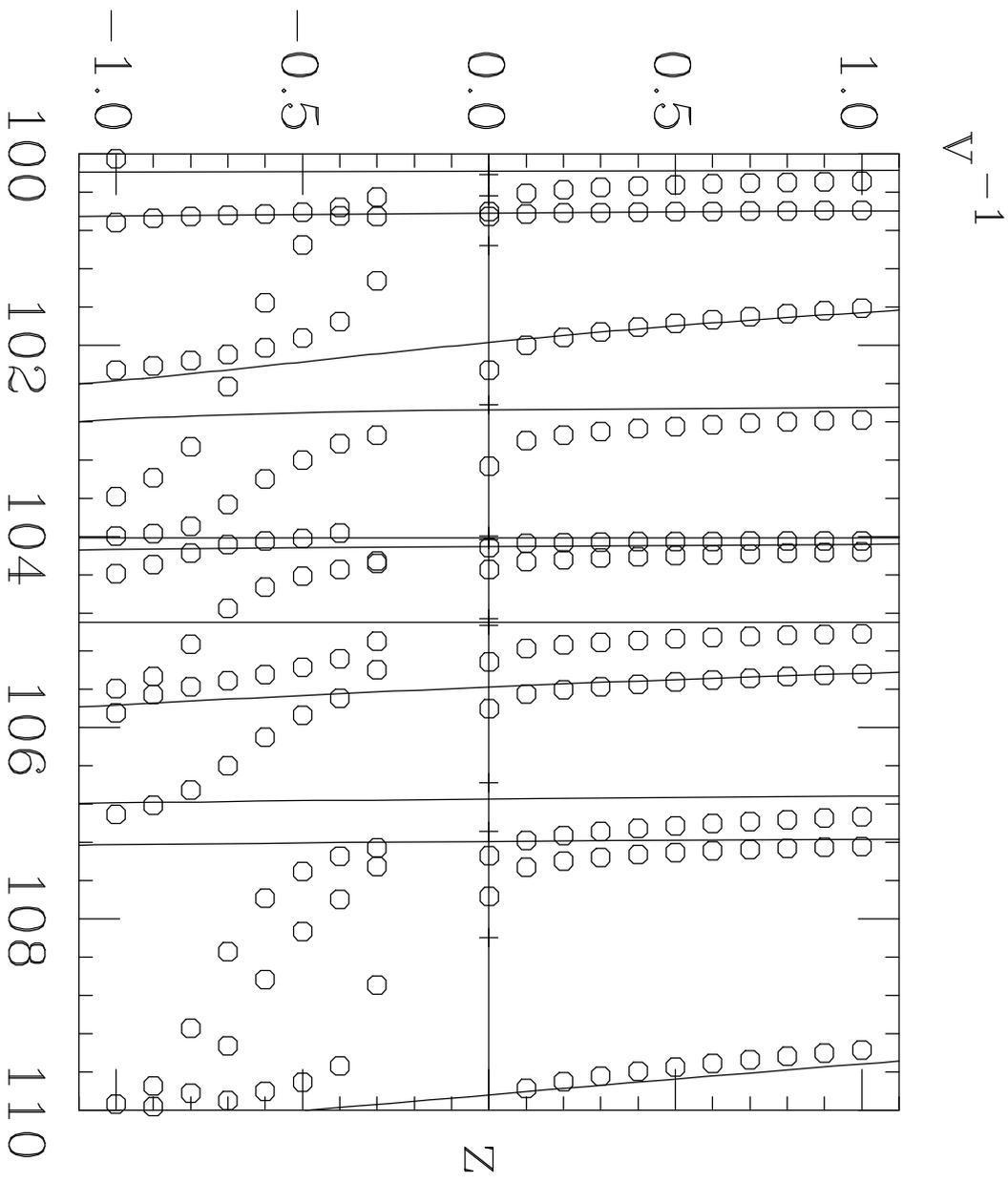
\begin{figure}
\caption{
The eigenvalues around  
$z \simeq  0.30 E_{N(\delta l_x,\delta l_y)}$.
The indications are the same as in Fig.1.  
}
\end{figure}
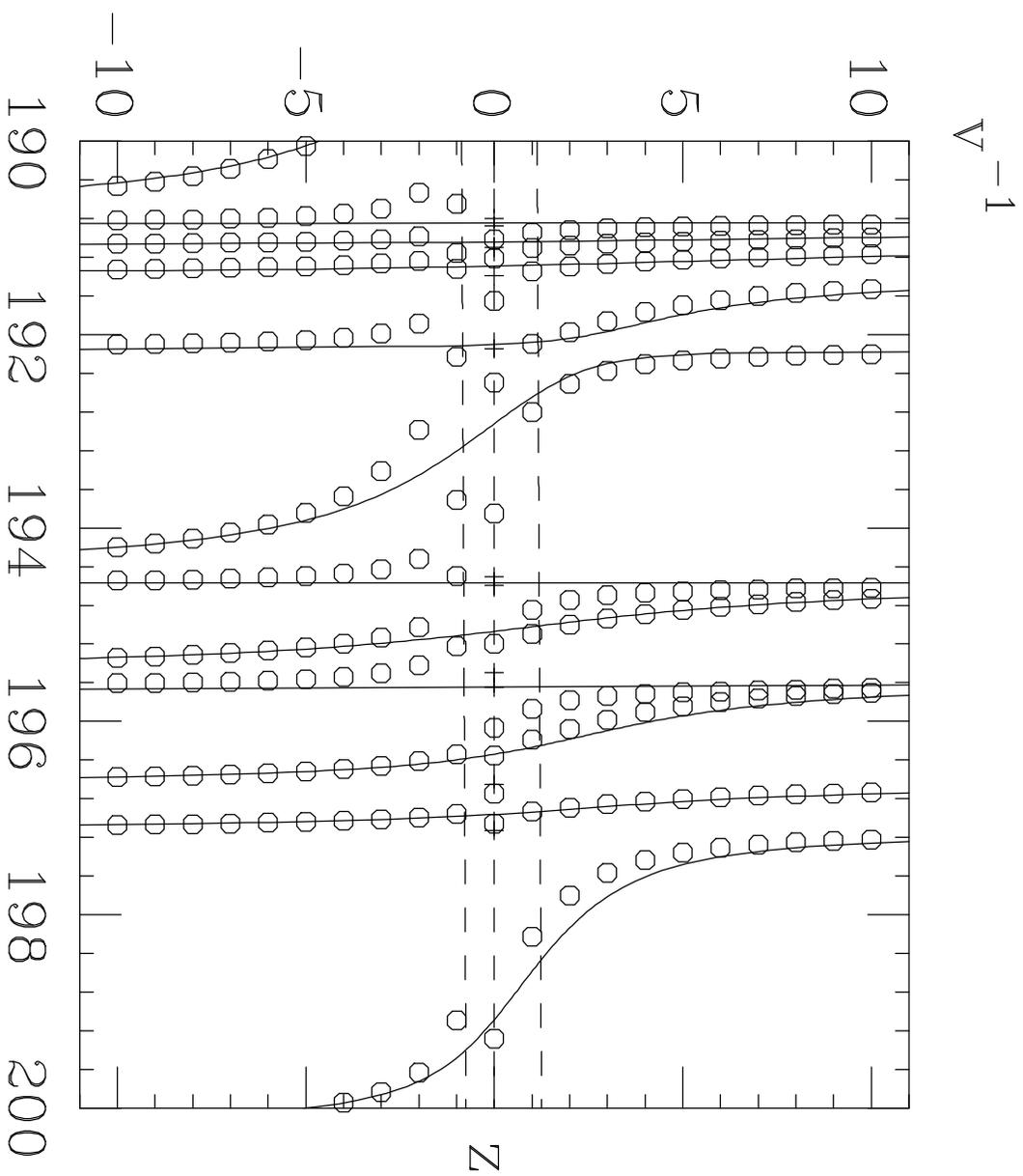
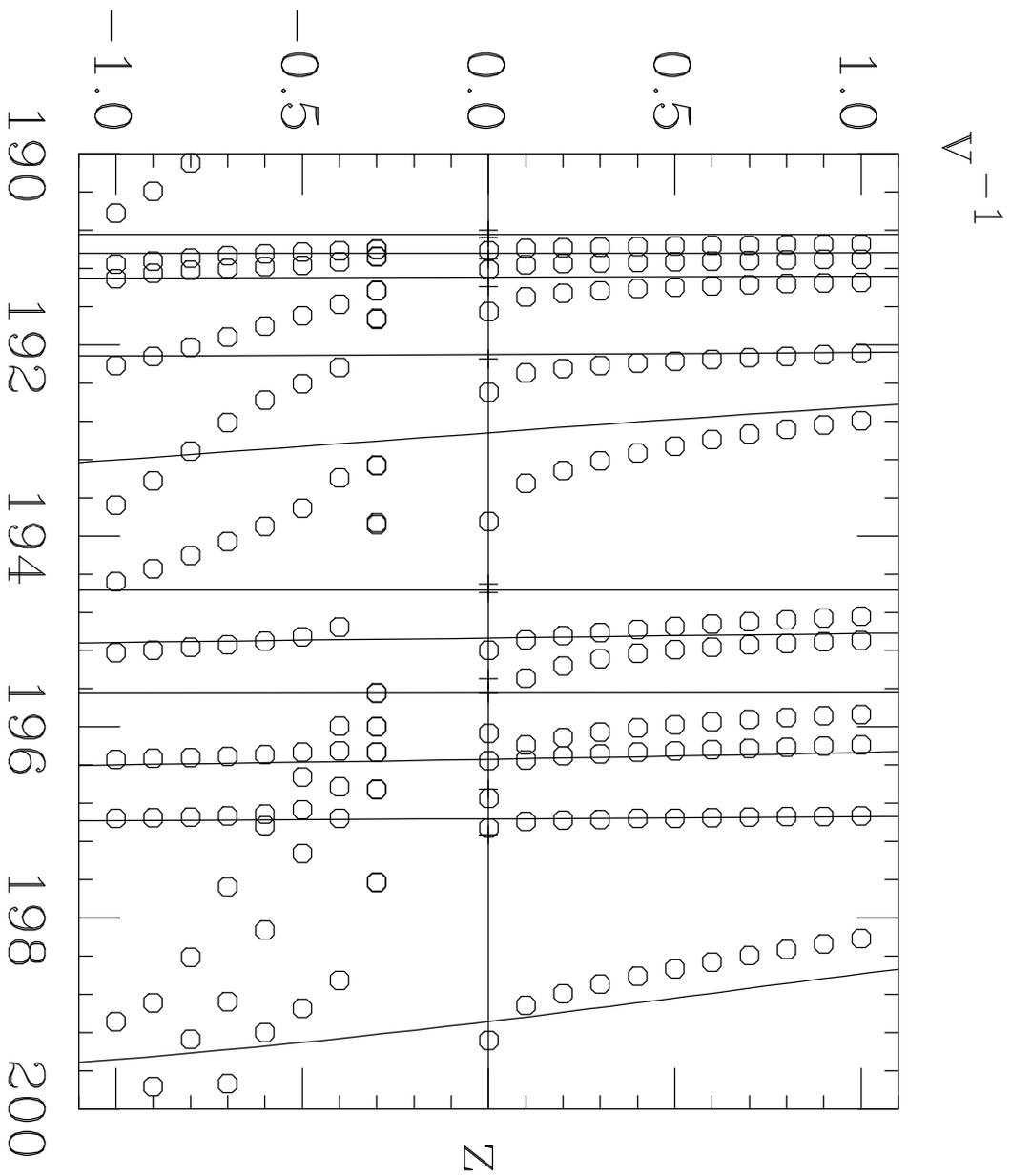
\begin{figure}
\caption{
The eigenvalues around  
$z \simeq  0.55 E_{N(\delta l_x,\delta l_y)}$.
The indications are the same as in Fig.1.  
}
\end{figure}
\end{document}